\documentclass[
    ,final            
  ]
  {aipproc}

\layoutstyle{6x9}

\begin{document}

\title{Prevalence of compact nuclear starbursts in nearby Seyfert galaxies}

\author{K. Kohno}{
  address={Institute of Astronomy, The University of Tokyo, Osawa, Mitaka, Tokyo, 181-0015, Japan}
}

\begin{abstract}
We present an imaging survey of the CO(1--0), HCN(1--0), and HCO$^+$(1--0) lines 
in the nearby Seyfert galaxies using the Nobeyama Millimeter Array and RAINBOW Interferometer.
Some of the observed Seyfert galaxies including NGC 1068, NGC 1097, NGC 5033, and NGC 5194 
exhibit strong HCN(1--0) emission on a few 100 pc scales. 
The observed HCN(1--0)/CO(1--0) and HCN(1--0)/HCO$^+$(1--0) line ratios 
in the Seyfert nuclei ($>$0.2 and $>$1.8, respectively) have never been observed 
in the central regions of nuclear starburst galaxies. 
On the other hand, the molecular line ratios in the nuclei of 
NGC 3079, NGC 3227, NGC 4051, NGC 6764, NGC 7479, and NGC 7469 are 
comparable with those in the nuclear starburst galaxies. 
We propose that the elevated HCN emission originates from the 
X-ray irradiated dense molecular tori or 
X-ray dominated regions (XDRs) in close proximity to the active nuclei. 
Further, our HCN/CO and HCN/HCO$^+$ diagrams will provide a new powerful diagnostic 
of the nuclear power source in active galaxies. Based on our diagnostic,
we observe three of five type-1 Seyferts (six of ten in total) host compact nuclear starbursts.
Our results are also supported by observations at other wavelengths
such as those by infrared L-band PAH spectroscopy. 
The proposed method based on the HCN and HCO$^+$ spectroscopies will be crucial for investigating
extremely dusty nuclei, such as the ultraluminous infrared galaxies and 
high-redshift submillimeter galaxies, 
because these molecular lines are devoid of dust extinction. 
As an example, we present the HCN and HCO$^+$ observations
of the luminous infrared galaxy NGC 4418. We have obtained a high HCN/HCO$^+$ ratio of 1.8,
which suggests the presence of a buried active nucleus.
\end{abstract}

\maketitle


\section{Survey of Dense Molecular Gas in Seyfert Galaxies using Nobeyama Millimeter Array}

A dense molecular medium plays various roles in the
vicinity of active galactic nuclei (AGNs). The presence of
dense and dusty interstellar matter (ISM), which obscures the
broad line regions in the AGNs, is inevitable at $<$  1pc --- a few 10 pc
scales according to the proposed unified model of Seyfert
galaxies. This circumnuclear dense
ISM could be a reservoir of fuel for the active nuclei and a site for 
massive star formation. In fact, strong HCN (1--0)
emission, which requires dense ($n_{\rm H_2} > 10^4$ cm$^{-3}$) environments
for its collisional excitation, has been detected in the prototypical
type-2 Seyfert NGC 1068 \citep{Jackson:1993,Tacconi:1994,Helfer:1995}.
Similar enhancements in the low-luminosity
Seyfert galaxies such as NGC 1097 \citep{Kohno:2003a}
and NGC 5194 \citep{Kohno:1996} have been reported.
In these Seyfert nuclei,the HCN (1--0) to CO (1--0) integrated intensity ratios in the 
brightness temperature scale, $R_{\rm HCN/CO}$, are enhanced up to approximately 0.4-0.6, 
and the kinematics of the HCN line indicates that this dense molecular medium could be the outer envelope of the
abovementioned obscuring material \citep{Jackson:1993,Tacconi:1994,Kohno:1996}. 

Howevwer, not every Seyfert galaxy exhibits such HCN enhancement. For example, 
there is no significant enhancement in $R_{\rm HCN/CO}$
within the central region extending over a few 100 pc of the Seyfert galaxy 
NGC 6951 \citep{Kohno:1999},
although a slight increase in the ratio can be observed in the circumnuclear starburst/star-forming
rings of NGC 6951. Questions regarding the difference between these Seyfert galaxies with and without
HCN enhancement and the nature
of the strong HCN emission toward some of Seyfert galaxies arise.

In order to address these issues, we have conducted an extensive ``3D'' imaging survey of
the CO(1--0), HCN(1--0), and HCO$^+$(1--0) lines in the local Seyfert galaxies 
using the Nobeyama Millimeter Array (NMA) and the RAINBOW interferometer
with typical resolutions of $\sim$ $2''$ to $6''$ and sensitivities of a few mJy beam$^{\rm -1}$
for a $\sim$ 50 km s$^{\rm -1}$ velocity channel.
The sample of galaxies for the survey is listed in Table~\ref{tab:a}. 
The majority of the galaxies are from the Palomar Northern Seyfert sample \citep{Ho:1997}.
Some southern Seyfert galaxies are also included in the sample. 
It should be noted that the HCN(1--0) and HCO$^+$(1--0) lines are observed simultaneously by using
the Ultra Wide Band Correlator \citep{Okumura:2000}. This is essential
to obtain accurate HCN/HCO$^+$ integrated intensity ratios ($R_{\rm HCN/HCO^+}$ hereafter).

\begin{table}
\begin{tabular}{ccccccc}
\hline
  \tablehead{1}{c}{b}{Name}
& \tablehead{1}{c}{b}{Class}
& \tablehead{1}{c}{b}{Morphology}
& \tablehead{1}{c}{b}{D\tablenote{Distance in Mpc.}}
& \tablehead{1}{c}{b}{CO\tablenote{The survey status. $\bigcirc$ = observed; 
$\triangle$ = observations in progress; blank = not observed.}}
& \tablehead{1}{c}{b}{HCN$^\dagger$}
& \tablehead{1}{c}{b}{HCO$^+$$^\dagger$} \\
\hline
\multicolumn{7}{c}{Seyfert galaxies listed in Palomar Northern sample}\\
\hline
NGC 1068 & S1.9 & (R)SA(rs)b & 14.4 &            & $\bigcirc$ & $\bigcirc$ \\
NGC 1667 & S2   & SAB(r)c    & 61.2 & $\bigcirc$ & $\bigcirc$ & $\bigcirc$ \\
NGC 3079 & S2   & SB(s)c     & 20.4 & $\bigcirc$ & $\bigcirc$ & $\bigcirc$ \\
NGC 3227 & S1.5 & SAB(s)a pec & 20.6 & $\bigcirc$ & $\bigcirc$ & $\bigcirc$ \\
NGC 3982 & S1.9 & SAB(r)b:   & 17.0 & $\bigcirc$ & $\bigcirc$ & $\bigcirc$ \\
NGC 4051 & S1.2 & SAB(rs)bc  & 17.0 & $\bigcirc$ & $\bigcirc$ & $\bigcirc$ \\
NGC 4151 & S1.5 & (R')SAB(rs)ab: & 20.3 & $\bigcirc$ &         &         \\
NGC 4258 & S1.9 & SAB(s)bc   & 6.8  & $\bigcirc$ & $\triangle$ & $\triangle$ \\
NGC 4388 & S2   & SA(s)b: sp & 16.8 & $\bigcirc$ & $\bigcirc$ & $\bigcirc$ \\
NGC 4501 & S2   & SA(rs)b    & 16.8 & $\bigcirc$    & $\bigcirc$ & $\bigcirc$ \\
NGC 4579 & S1.9/L1.9 & SAB(rs)b  & 16.8 & $\bigcirc$    &         &         \\
NGC 5033 & S1.5 & SA(s)c     & 18.7 & $\bigcirc$ & $\bigcirc$ & $\bigcirc$ \\
NGC 5194 & S2   & SA(s)bc pec & 7.7 & $\bigcirc$ & $\bigcirc$ & $\bigcirc$ \\
NGC 6951 & S2   & SAB(rs)bc  & 24.1 & $\bigcirc$ & $\bigcirc$ & $\triangle$ \\
NGC 7479 & S1.9 & SB(s)c     & 32.4 & $\bigcirc$ & $\bigcirc$ & $\bigcirc$ \\
\hline
\multicolumn{7}{c}{Southern Seyfert galaxies} \\
\hline
NGC 1097 & S1   & (R')SB(r'l)b & 14.0 & $\bigcirc$ & $\bigcirc$ & $\bigcirc$ \\
NGC 5135 & S2   & SB(l)ab    & 54.8 & $\bigcirc$ &            &         \\
NGC 6764 & S2   & SB(s)bc    & 32.2 & $\bigcirc$ & $\bigcirc$ & $\bigcirc$ \\
NGC 6814 & S1.5 & SAB(rs)bc  & 20.8 & $\bigcirc$ &  &  \\
NGC 7465 & S2   & (R')SB(s)0 & 26.2 & $\bigcirc$ & $\bigcirc$  & $\bigcirc$ \\
NGC 7469 & S1.2 & (R')SAB(rs)a & 65.2 & $\bigcirc$ & $\bigcirc$ & $\bigcirc$ \\
\hline
\end{tabular}
\caption{Galaxy sample for the survey using the Nobeyama Millimeter Array }
\label{tab:a}
\end{table}

\subsection{Results: type-1 Seyfert galaxies NGC 5033 and NGC 3227}

Here, the CO(1--0), HCN(1--0), and HCO$^+$(1--0) images of the two type-1 Seyfert galaxies 
NGC 5033 and NGC 3227 are shown in Figure 1 and 2, respectively. 

We observe a significant enhancement in the
HCN(1--0) emissions at the center of NGC 5033; the CO emission exhibits a twin peaks morphology
\citep{Kohno:2003b}, whereas HCN emission originates from the very center of the galaxy.
The HCO$^+$ emission also exhibits a single peak at the nucleus; however, 
the weakness of the nuclear HCO$^+$ emission as compared with the HCN emission is remarkable.
The $R_{\rm HCN/CO}$ of the nucleus of NGC 5033 is enhanced
up to 0.23, and the $R_{\rm HCN/HCO^+}$
is approximately equal to 2.
This is the fourth detection of the abnormally enhanced HCN emission
($R_{\rm HCN/CO} > 0.2$ and $R_{\rm HCN/HCO^+} > 1.8$) among the Seyfert galaxies
after NGC 1068, NGC 5194, and NGC 1097
\citep{Jackson:1993,Kohno:1996,Kohno:2003a}.

On the other hand, we observe moderate HCN(1--0) emission toward NGC 3227;
the observed values of $R_{\rm HCN/CO}$ and $R_{\rm HCN/HCO^+}$ are 0.043 and 0.83, respectively.

\begin{figure}
  \includegraphics[height=.25\textheight]{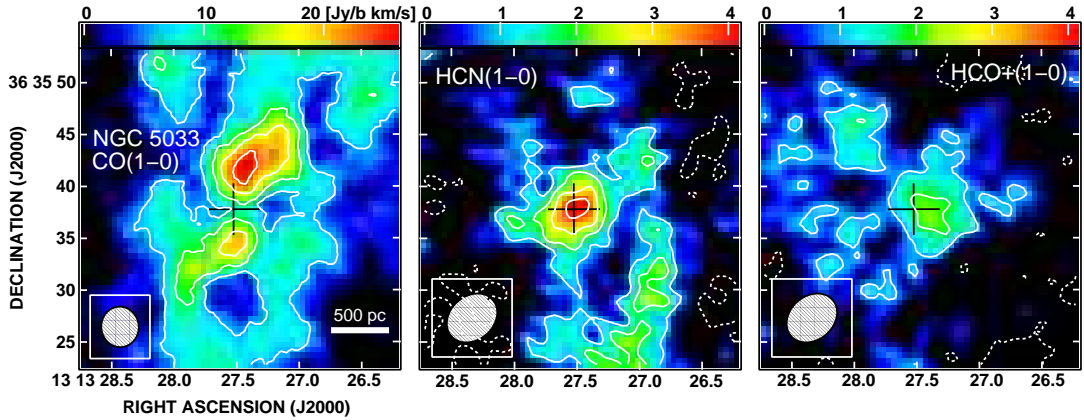}
  \caption{The integrated intensity images of the CO(1--0), HCN(1--0), and HCO$^+$(1--0) lines
in the central region extending over a few kpc of the type-1 Seyfert galaxy NGC 5033. 
There ia a strong concentration of the HCN(1--0) emission
toward the active nucleus; in contrast, the CO(1--0) image that exhibits the twin peaks morphology
with no significant source at the very center. The weakness of the HCO$^+$(1--0) emission is also remarkable.}
\end{figure}

\begin{figure}
  \includegraphics[height=.27\textheight]{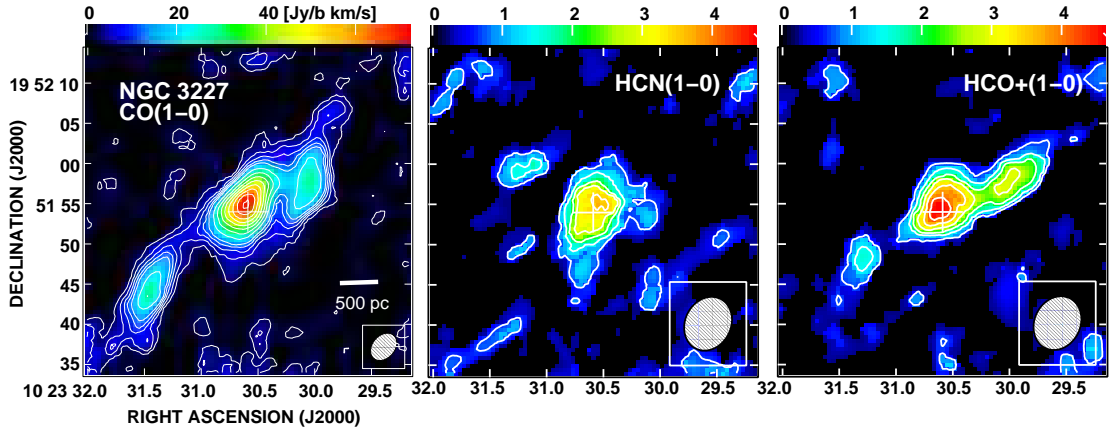}
  \caption{The integrated intensity images of the CO(1--0), HCN(1--0), and HCO$^+$(1--0) lines
in the central region extending over a few kpc of the type-1 Seyfert galaxy NGC 3227. 
No significant enhancement in the HCN emission is observed.}
\end{figure}

\section{HCN/HCO$^+$ Ratios as a New Diagnostic of Nuclear Power Sources in Active Galaxies}

Figure 3 compares the observed line ratios of the Seyfert galaxies
with those of the nuclear starburst galaxies,
which were also measured with similar angular resolutions using the Nobeyama 
Millimeter Array \citep{Kohno:2001}.

Some of the observed Seyfert galaxies including NGC 1068, NGC 1097, NGC 5033, and NGC 5194 
exhibit strong HCN(1--0) emission on a few 100 pc scales; 
the observed $R_{\rm HCN/CO}$ and $R_{\rm HCN/HCO^+}$ values of 
these Seyfert nuclei ($>$0.2 and $>$1.8, respectively) heve never been observed 
in the central regions of the nuclear starburst galaxies. 
On the other hand, the molecular line ratios in the nuclei of 
NGC 3079, NGC 3227, NGC 4051, NGC 6764, NGC 7479, and NGC 7469 are 
comparable with those in the nuclear starburst galaxies. 
It should be noted that very high $R_{\rm HCN/HCO^+}$ values
in NGC 3079 and Maffei 2 ($>3$) were reported previously
\citep{Nguyen:1992}; however, our new simultaneous measurements yielded
moderate ($\sim 1$) ratios.

We propose that the two groups in our ``HCN diagram''
(Figure 3) can be explained 
in terms of the ``AGN - nuclear starburst connection''
\footnote{Note that this differs from the ``AGN - circumnuclear starburst
cohabitation,'' which often refers to the association
of the AGN with star formation on galactic disk scales in the AGN hosts}.

In the Seyferts whose line ratios are comparable to those
in the nuclear starburst galaxies, it seems likely
that a compact nuclear starburst,
which presumably exists in the dense molecular torus
\citep{wada:2002},
is associated with the Seyfert
nucleus (i.e., ``composite'').
In the nuclear regions of composite Seyferts,
the fractional abundance of HCO$^+$ is expected to 
increase due to frequent supernova (SN) explosions.
In fact, in evolved starbursts such as M82, 
in which large-scale outflows
have occurred due to numerous
SN explosions, 
HCO$^+$ is often stronger than HCN
\citep{Nguyen:1992}.

On the other hand, 
the HCN-enhanced Seyferts, with
$R_{\rm HCN/CO} > 0.2$ and $R_{\rm HCN/HCO^+} > 1.8$,
would host ``pure'' AGNs, 
with the absence of any associated nuclear starburst activity. 
We propose that the elevated HCN emission originates from 
X-ray irradiated dense molecular tori or 
X-ray dominated regions (XDRs) \citep{Maloney:1996} in close proximity to the active nuclei 
because it has been predicted that the fractional abundance of
HCN is enhanced by the strong X-ray radiation
from the AGN \citep{Lepp:1996}, thereby resulting 
in abnormally high $R_{\rm HCN/CO}$ and $R_{\rm HCN/HCO^+}$
values. 
In fact, based on a molecular line survey \citep{Usero:2004}, 
it is proposed that the circumnuclear molecular disk
at the center of NGC 1068 is a giant XDR.

\begin{figure}
  \includegraphics[height=.37\textheight]{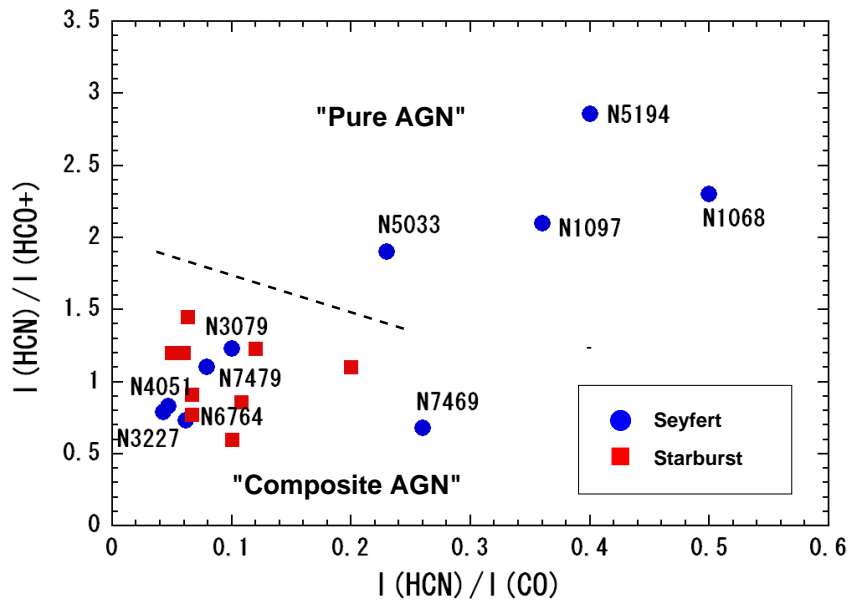}
  \caption{HCN/CO and HCN/HCO$^+$ integrated intensity ratios of the Seyfert and starburst galaxies. 
Some Seyfert galaxies, such as NGC 1068, NGC 1097, NGC 5033, and NGC 5194, exhibit extremely enhanced
HCN(1--0) emission with respect to CO(1--0) and HCO$^+$(1--0) lines, 
whereas other Seyfert galaxies exhibit line ratios that are very similar to 
those in the nuclear starburst galaxies.}
\end{figure}

\subsection{Comparison with another diagnostic: L-band PAH spectroscopy}

Our interpretation is also supported by other wavelength data.
In this section, we compare our results with those of infrared L-band spectroscopy surveys
\citep{Imanishi:2002,Rodriguez:2003}.
A polycyclic aromatic hydrocarbon (PAH) emission at 3.3 $\mu$m
in the L band
is commonly observed in the starburst regions; however, it is easily destroyed in 
the vicinity of the active nuclei due to their hard radiation field. 
Thus, L-band spectroscopy can be considered as a powerful tool to study
nuclear starbursts in Seyfert galaxies.

No strong PAH emission in the center of NGC 5033 has been reported \citep{Imanishi:2002}.
This is consistent with our detection of enhanced HCN emission toward the center.
In other words, NGC 5033 hosts a pure Seyfert nucleus, suggesting that the properties of the circumnuclear
molecular gas in the vicinity of NGC 5033 are governed by XDR chemistry.
In contrast to NGC 5033, NGC 3227 exhibits 
intense PAH emission feature at 3.3 $\mu$m \citep{Imanishi:2002,Rodriguez:2003},
which indicates the presence of a compact nuclear starburst 
around the type-1 Seyfert nucleus of NGC 3227.
This is consistent with moderate $R_{\rm HCN/CO}$ and $R_{\rm HCN/HCO^+}$ values at the center of NGC 3227.

\section{Application to LIRGs: A Case for NGC 4418}

The HCN/CO and HCN/HCO+ diagrams shown in Figure 3 will provide us with a powerful diagnostic of the power source 
in the extremely dusty nuclei such as the ultraluminous infrared galaxies (ULIRGs) and 
high-redshift dusty starbursts, such as the submillimeter galaxies, 
because these molecular lines are devoid of any dust extinction. 

Recently, we have conducted high resolution HCN and HCO$^+$ observations 
of the LIRG NGC 4418 using the RAINBOW interferometer. 
We detected compact and enhanced HCN emission (the observed 
$R_{\rm HCN/HCO^+}$ was 1.8), which indicates the presence of
a buried active nucleus at the center of NGC 4418 \citep{Imanishi:2004}.


\begin{theacknowledgments}
I thank my collaborators including T.\ Shibatsuka, M.\ Okiura, K.\ Nakanishi,
T.\ Tosaki, T.\ Okuda, S.\ Onodera, M.\ Doi, K.\ Muraoka, A.\ Endo, 
S.\ Ishizuki, K.\ Sorai, S.\ K.\ Okumura, Y.\ Sofue,
R.\ Kawabe, B.\ Vila-Vilar\'o, and M.\ Imanishi 
for their efforts to make this survey possible.
We are grateful to the NRO/NMA staff for the operation and 
continuous efforts to improve the NMA.
This study was financially supported by
the JSPS Grant-in-Aid for Scientific Research (B) No.14403001 and 
MEXT Grant-in-Aid for Scientific Research on Priority Areas No.15071202. 
\end{theacknowledgments}


\bibliographystyle{aipproc}   

\bibliography{kohno}

\IfFileExists{\jobname.bbl}{}
 {\typeout{}
  \typeout{******************************************}
  \typeout{** Please run "bibtex \jobname" to obtain}
  \typeout{** the bibliography and then re-run LaTeX}
  \typeout{** twice to fix the references!}
  \typeout{******************************************}
  \typeout{}
 }

\end{document}